\documentstyle[12pt,cite,epsfig]{article}

\newcommand{\be}{\begin{equation}}
\newcommand{\ee}{\end{equation}}
\newcommand{\bea}{\begin{eqnarray}}
\newcommand{\eea}{\end{eqnarray}}

\def\half{\frac{1}{2}}

\def\IB{\relax\hbox{$\inbar\kern-.3em{\rm B}$}}
\def\IC{\relax\hbox{$\inbar\kern-.3em{\rm C}$}}
\def\ID{\relax\hbox{$\inbar\kern-.3em{\rm D}$}}
\def\IE{\relax\hbox{$\inbar\kern-.3em{\rm E}$}}
\def\IF{\relax\hbox{$\inbar\kern-.3em{\rm F}$}}
\def\IG{\relax\hbox{$\inbar\kern-.3em{\rm G}$}}
\def\IGa{\relax\hbox{${\rm I}\kern-.18em\Gamma$}}
\def\IH{\relax{\rm I\kern-.18em H}}
\def\IK{\relax{\rm I\kern-.18em K}}
\def\IL{\relax{\rm I\kern-.18em L}}
\def\IP{\relax{\rm I\kern-.18em P}}
\def\IR{\relax{\rm I\kern-.18em R}}
\def\IZ{\relax{\rm Z\kern-.5em Z}}

\begin{document}

\begin{titlepage}

\begin{flushright}
SNUTP/05-003\\
hep-th/0503164\\
\end{flushright}

\vskip .5in

\begin{center}

{\Large\bf Note on Brane-antibrane Description of Non-extremal Black Holes}

\vskip .5in

{\bf Sanjay Siwach}\footnote{e-mail: {sksiwach@phya.snu.ac.kr}},

\vskip .2in

{\it Centre for Theoretical Physics,\\
 Seoul National University,
 Seoul - 151 742, Korea }\\

and

{\it Physics Department,\\
 University of Seoul,
 Seoul - 130 743, Korea }\\

\vspace{.3in}

\begin{abstract}

\vskip .2in

Recently a number of authors have used a system of branes and antibranes at finite temperature for microscopic modeling of non-extremal black holes in string theory. The entropy of the system is derived from the simplified assumption of decoupled gas of open strings on the stacks of branes and antibranes and extremizing the total entropy with respect to the number of branes (or antibranes). The resulting entropy differs from the supergravity entropy by a deficit factor. We give an intuitive explanation for the deficit factor. We treat the whole system as two stacks of branes and antibranes with a single copy of Yang-Mills gas common to both the stacks. This gives the answer in agreement with the supergravity.

\end{abstract}

\end{center}

\vfill

\end{titlepage}


Black holes have remained mysterious objects since their birth in General Theory of Relativity. However some breakthrough has been achieved in last 30-40 years in unraveling a part of the mystery. First one was the Bekenstein's observation (motivated by area theorems) that the area of event horizon can be interpreted as the entropy of the black hole and the discovery of Hawking radiation and subsequent formulation of the laws of black hole thermodynamics. (Please see \cite{dam,wald} for recent reviews about these developments.) The second one being the microscopic calculation of entropy by Strominger and Vafa for certain black holes in string theory \cite{sv}. 

In string theory, black holes are the solutions of low energy effective actions and are also known as black p-branes (for a review, please see \cite{youm}). The counting of Dp-brane microstates by Strominger and Vafa was the first hint that Dp-branes can provide the microscopic theory of black holes. Dp-brane dynamics (open string gas on the worldvolume) successfully accounts the microscopic entropy of the near extremal black holes in string theory. Low energy degrees of freedom on Dp-brane worldvolume are the massless modes of super Yang-Mills on the worldvolume of the Dp-branes. This also suggests that gravitational degrees of freedom of string theory black holes at low energy can be described by field theoretic Yang-Mills degrees of freedom and perhaps in a holographic manner. Despite these efforts a satisfactory resolution of black hole information loss paradox is still lacking (for a brief exposition of current state of affairs, please see \cite{russo}).

To understand the black holes far from extremality and Schwarzschild black holes, it appears that both branes and antibranes are required \cite{hms,dgk}. A brane-antibrane system is unstable and decays via tachyon condensation. But at high enough temperatures the system can exist in a thermodynamically stable state. The entropy of non-extremal (and Schwarzschild) black holes can be reproduced by such a system upto a deficit factor \cite {dgk}. The negative specific heat of the charged and the Schwarzschild black holes can be understood from the dynamics of brane-antibrane pair annihilation. This work has been generalized to charged, multicharged and rotating configurations in subsequent works \cite{ghm,peet,bl,krama,l,ks,halyo} (see also the related works \cite{l1, gg}). 

To describe the non-extremal black holes one takes a stack of $N$ branes and a stack of $\bar N$ antibranes. The total energy of the system can be thought of as sum of extremal energy due to branes and antibranes and the energy above extremality due to open string excitations. One assumes that the gas of open strings on branes does not interact with the gas on antibranes and then the entropy of the system is just the sum of entropies due to contributions from branes and antibranes. One further assumes that energy due to open string excitations on the branes is equal to the one on the antibranes and extremizes the entropy with respect to the number of branes (or antibranes), while keeping the total energy and charge fixed. The resulting entropy agrees with the supergravity entropy in functional form but differs by a numerical factor. 
 
Here we attempt to give a plausible explanation for the deficit factor. We shall treat the full system as two stacks of branes and antibranes as before but with a single copy of the Yang-Mills gas sourced by both the stacks of branes and antibranes. We propose an expression for the entropy of the system in terms of energy density of the Yang-Mills gas. This is motivated by the key observation that the entropy of the system depends explicitly on the energy density rather than energy on the stacks of branes or antibranes. Now extremizing the entropy of the system with respect to number of branes (or antibranes), one gets the answer in perfect agreement with the supergravity. The explanation is in the spirit of \cite{krama} (see also \cite{ks}) but actual details are different. We shall restrict ourselves to the discussion of charged black holes following \cite{dgk,l,ks}. 

We are interested in microscopic modeling of non-extremal black holes in terms of a system of branes and antibranes at finite temperature. As mentioned earlier, at finite temperature the unstable vacuum of the tachyon can become a stable minimum of tachyon potential and the system can exist in a thermodynamically stable state characterized by minimum of the free energy. To derive the thermodynamics of the system one can consider a stack of $N$ coincident branes and a stack of $\bar N$ coincident antibranes. The entropy due to massless degrees of freedom on the brane (antibrane) worldvolume can be derived from the near extremal black holes in supergravity \cite{kt}. 

Consider $p$-brane solutions of supergravity in $D = d + p + 3$ dimensional spacetime. Wrapping $p$ spatial dimensions on a $p$-torus $T^p$, we get a charged black hole in $D - p = d + 3$ dimensional spacetime. 

The ADM mass and the entropy of the these black holes is given by:
\be
\label{sgentropy}
M_0  =  b \mu (\lambda + \half \cosh 2 \phi), \hspace{0.5cm}
S_0  =  c \mu^{\lambda + \half}~\cosh \phi.
\ee

The constants $b$ and $c$ involve the volume factors etc. and can be read from the references \cite{kt,ct} and $\lambda$ is defined as:
\[
\lambda = \frac{d + 1}{d} - \half.
\]

The extremal limit corresponds to taking $\mu \rightarrow 0$ and $\phi \rightarrow \infty$ while keeping, $Q = \frac{\mu}{2} \sinh {2\phi}$ ~ fixed. From the above equations one can write the ADM mass and the entropy for the stack of $N$ coincident branes as:
\be
M = b N  + E , \hspace{0.5cm} S = c_1 \sqrt{N} E^\lambda
\ee
where $E$, is the energy above extremality and the constant $c_1$ is proportional to $c$ (modulo factors of $2\pi$ etc.). 

We shall denote the corresponding expression for antibranes with a bar. Further we shall restrict ourselves to $\lambda > 0$ in this note, so that the extremal entropy is zero. Note that the most interesting examples of three charge black hole in 5-dimensions and four charge black holes in 4-dimensions have non-zero extremal entropy and hence corresponds to $\lambda = 0$. The discussion of black holes with non-zero extremal entropy is more involved and no satisfactory treatment exists for these black holes in the present framework (for some attempts, please see \cite{bl}).

Now let us consider the system of $N$ branes and $\bar N$ antibranes with charge, $Q = N - \bar N$. To describe the thermodynamics of the system one assumes that thermodynamics of stack of branes is decoupled from the thermodynamics of antibranes \cite{dgk}. The expression for ADM mass can be written as: 
\be
M = b (N + \bar{N}) + E  + \bar{E}.
\ee

Given the assumption of decoupled thermodynamics for the stack of branes and antibranes, the total entropy of the system is just the sum due to individual contribution:
\be
\label{ftentropy}
S = c_1 \left(\sqrt{N} E^\lambda +  \sqrt{\bar{N}} \bar{E}^\lambda  \right). 
\ee

For fixed $N$ or $\bar N$, the entropy is given by this expression. However in dynamical situations, brane-antibrane pairs create and annihilate constantly and  the entropy of the system is determined by extremising the entropy (\ref{ftentropy}) with respect to number of branes $N$ (or antibranes $\bar N$) while keeping the total energy, $M$ and the charge, $Q$ fixed. One further assumes that $E = \bar{E}$. Extremization of the entropy (\ref{ftentropy}) gives the following equation for $N$ (or $\bar N$):
\be
M - b (N + \bar{N}) = 4 \lambda b \sqrt{N \bar{N}}.
\ee

For $N = \bar N$, the solution is easy and $N \propto M$. For $N \ne \bar N$, the solution can be written in the parametric form:
\[
N = \frac{m}{2} e^{2 \theta}, \hspace{0.5cm}
\bar{N} = \frac{m}{2} e^{- 2 \theta} .
\]

To compare with supergravity entropy (\ref{sgentropy}), one has to equate $m = \mu$ and $\theta = \phi$ so that $M = M_0$ and $Q = Q_0$ and we get,

\be
\label{cf}
S (M, Q) = 2^{- \lambda} S_0 (M, Q)  
\ee

that is the field theory entropy is smaller by a factor of $2^{- \lambda}$ compared to the supergravity entropy. 

We shall see below that the mismatch has to do with the assumption of decoupled thermodynamics for the Yang-Mills gases on the stacks of branes and antibranes. We shall give up this assumption and treat the whole system as two stacks of branes and antibranes as before but with a single copy of Yang-Mills gas common to them. In other words, the Yang-Mills gases on the two stacks are no longer decoupled but interact to make a single copy common to both the stacks. We shall see that with this new input, the field theory entropy is in agreement with supergravity answer. Also we are assuming that open strings stretched between brane and antibrane are heavy at low energies as usual

Since we are dealing with interacting Yang-Mills here, we need an expression for the entropy of the system (the entropy of the system will not be just the sum of contributions due to branes and antibranes because of the interactions). We shall propose an expression for the entropy below which captures the thermodynamics of the system. The key observation is that it is the energy density that appears explicitly in the entropy expression (\ref{ftentropy}) rather than the energy on the stacks of branes or antibranes. Note that the constant $c_1$ appearing in the entropy expression (\ref{ftentropy}) involves the factor of $V^{-\lambda}$ in addition to usual one factor of volume, $V$ and the volume in question is the volume of p-torus on which the Dp-branes are wrapped. 

Let us rewrite the entropy expression for the stack of branes in terms of energy density: 
\be
S= d_1 \sqrt N e^\lambda.
\ee

the constant $d_1$ is defined as $d_1 = c_1 V^{\lambda}$ and involves one factor of volume only. In the absence of antibranes the energy density, $e$ receives contribution from the open string gas on the branes only. When both branes and antibranes are present, the energy density also receives contribution from open string gas on antibranes. Roughly the energy density is double (for equal number of branes and antibranes) due to this contribution. 

The expression for ADM mass of the system can be written as:
\be
\label{adm}
M =  b ( N + \bar N)  + V e.
\ee

We propose the following expression for the entropy of the brane-antibrane system:
\be
\label{newentropy}
S = d_1 ( \sqrt{N} + \sqrt{\bar N}) e^\lambda
\ee
where $e$ is the total energy density of the open string gas due to both branes and antibranes, the one appearing in equation (\ref{adm}). Please note that this expression for entropy is quite different from the one (\ref{ftentropy}) written in terms of  energy, $E$ (because $e$ is total energy density now i.e. $e = 2E / V$).  At this stage, no first principle derivation of the expression for the entropy (\ref{newentropy}) exists from the dynamics of brane-antibrane system, but we give the following arguments in favour of the proposal.

1. One can do the extremization of the entropy (\ref{newentropy}) with respect to number of branes $N$ (or antibranes $\bar N$), while keeping the total energy, $M$ and charge, $Q$ fixed as before. The resulting answer is in perfect agreement with the supergravity answer.

2. For fixed number of branes and antibranes, we expect from the above discussion that,  $S\propto e^\lambda$. In the variational approach, when we allow the numbers of branes (or antibranes) to vary and extremize the entropy, we expect the $e^\lambda$ to be multiplied by square root of $N$ or $\bar N$ in order to reproduce the correct dependence of entropy with energy for a non-extremal black hole. In this sense we can think of the expression (\ref{newentropy}) as motivated by supergravity. 

3. In the near extremal limit $N$ (or $\bar N$) is zero, the expression (\ref{newentropy}) reduces to the standard expression for branes (or antibranes) derived from the supergravity. The co-efficient $d_1$ is fixed by the fact that in this limit, the branes (antibranes) describe the dynamics of near extremal black hole and can be read from the corresponding supergravity expression for entropy.

4. One can also think of some other candidate expression for the entropy of the system which reduces to the expression for branes (antibranes) in the extremal limit. For example, let us consider the following expression: 

\be
\label{new1}
S = d_1 \left(\frac{N + \bar N}{ \sqrt{N} + \sqrt{\bar N}}\right) e^\lambda.
\ee

Though the above expression gives correct near-extremal limit, but it means that branes and antibranes interact in a complicated manner to mix the stacks of branes and antibranes, in contradiction with our assumption that we treat the system as two stacks of branes and antibranes. Also this will give the answer in conflict with supergravity (if one can do the extremization at all). 

5. As in the above example, any explicit dependence of contribution of one stack of branes (antibranes) to the total entropy on the number of antibranes (branes) on the other stack will be problematic. So we are left with the only choice of the expression for the entropy (\ref{newentropy}), which has the correct near extremal limit. 

Now let us compare our proposal for the entropy expression (\ref{newentropy}) with the scaling arguments of the references \cite{krama,ks}. The scaling arguments are as follows, if one normalizes the brane energy (or brane tension), energy for each copy of the gas and the entropy of the system by the constant factors, say $\alpha$, $\epsilon $ and  $\sigma$ respectively, what one finds that the deficit factor $2^{-\lambda}$ appearing in the equation (\ref{cf}) can be made equal to unity if $\alpha = 1/4$, $\epsilon = 2$ and $\sigma = \half$ i.e. as if the total energy of the system is taken by a single copy of the gas and brane tension is reduced by one fourth. Our proposal for the entropy is similar to this. We have taken the system as a whole or just one copy of the open string gas on the stack of branes and antibranes. Other way to make the deficit factor equal to unity is by taking $\alpha = 1$, $\epsilon = 2 $ and $\sigma = 1$ i.e. as if each copy of the gas carries twice the energy available to it, but this violates the energy conservation. What we have emphasized in this note is that what matters is the total energy density rather than the energy on the stack of branes or antibranes, in the entropy expression and hence we don't have to assign twice the available energy to each copy of the gas. Moreover no scaling of energy or entropy etc. is required.

In summary, we have given an explanation for the deficit factor based on the fact that we can treat the system as two stacks of branes and antibranes and a single copy of Yang-Mills gas common to them. We observed that it is the total energy density rather than energy on the stack of branes or antibranes, that should appear in the entropy formula explicitly. Using this fact, we proposed a candidate expression for the entropy of the system which gives the agreement with the supergravity answer. Moreover the proposed expression seems to resolve the conceptual difficulties felt by earlier workers in an attempt to explain the discrepancy between supergravity and field theory answers.  

Here we have discussed single charged black holes only. The discussion can be easily generalized to multicharged black holes (and Schwarzschild black holes). The understanding of the near extremal limit of the system and transition from negative specific heat to positive specific heat for charged black holes are some remaining issues to be studied in more detail. Also the first principle derivation of entropy expression (\ref{newentropy}) from the knowledge of brane-antibrane system at finite temperature is still lacking. This may require the use of the techniques of string field theory. We wish to return to one or more of these issues in near future.

\vspace{5mm}

\noindent
{\bf Acknowledgments:} I would like to thank S. Kalyana Rama for useful correspondence and critical comments on the manuscript of the paper. This work is supported by KRF, Prof. J. E. Kim grant No. R14-2003-012-01001-0. 



\begin{thebibliography}{999}

\bibitem{dam}
 T.~Damour,
[arXiv:hep-th/0401160].

\bibitem{wald}
 R.~M.~Wald,
 Living Rev.\ Rel.\  {\bf 4} (2001) 6
 [arXiv:gr-qc/9912119].

\bibitem{sv}
 A.~Strominger and C.~Vafa,
 Phys.\ Lett.\ B {\bf 379} (1996) 99
 [arXiv:hep-th/9601029].


\bibitem{youm}
 D.~Youm,
 Phys.\ Rept.\  {\bf 316} (1999) 1
 [arXiv:hep-th/9710046].

\bibitem{russo}
J.~G.~Russo,
[arXiv:hep-th/0501132].

\bibitem{hms}
 G.~T.~Horowitz, J.~M.~Maldacena and A.~Strominger,
 Phys.\ Lett.\ B {\bf 383} (1996) 151
 [arXiv:hep-th/9603109].


\bibitem{dgk}
U.~H.~Danielsson, A.~Guijosa and M.~Kruczenski,
JHEP {\bf 0109} (2001) 011
[arXiv:hep-th/0106201].
Rev.\ Mex.\ Fis.\  {\bf 49S2} (2003) 61
[arXiv:gr-qc/0204010].


\bibitem{ghm}
A.~Guijosa, H.~H.~Hernandez Hernandez and H.~A.~Morales Tecotl,
JHEP {\bf 0403} (2004) 069
[arXiv:hep-th/0402158].

\bibitem{peet}
O.~Saremi and A.~W.~Peet,
Phys.\ Rev.\ D {\bf 70} (2004) 026008
[arXiv:hep-th/0403170].

\bibitem{bl}
O.~Bergman and G.~Lifschytz,
JHEP {\bf 0404} (2004) 060
[arXiv:hep-th/0403189].

\bibitem{krama}
 S.~Kalyana Rama,
 Phys.\ Lett.\ B {\bf 593} (2004) 227
 [arXiv:hep-th/0404026].

\bibitem{l}
 G.~Lifschytz,
 JHEP {\bf 0409} (2004) 009
 [arXiv:hep-th/0405042].

\bibitem{ks}
 S.~K.~Rama and S.~Siwach,
 Phys.\ Lett.\ B {\bf 596} (2004) 221
 [arXiv:hep-th/0405084].

\bibitem{halyo}
 E.~Halyo,
 JHEP {\bf 0410} (2004) 007
 [arXiv:hep-th/0406082].

\bibitem{l1}
G.~Lifschytz,
JHEP {\bf 0408} (2004) 059
[arXiv:hep-th/0406203].

\bibitem{gg}
J.~A.~Garcia and A.~Guijosa,
JHEP {\bf 0409} (2004) 027
[arXiv:hep-th/0407075].

\bibitem{kt}
I.~R.~Klebanov and A.~A.~Tseytlin,
Nucl.\ Phys.\ B {\bf 475} (1996) 164
[arXiv:hep-th/9604089].

\bibitem{ct}
 M.~Cvetic and A.~A.~Tseytlin,
 Nucl.\ Phys.\ B {\bf 478} (1996) 181
 [arXiv:hep-th/9606033].



\end{thebibliography}
\end{document}